# Electron-phonon mediated superconductivity in $La_6Ni_5O_{12}$ nickel oxides

Alvaro Adrian Carrasco Alvarez,[1,2] Sébastien Petit,[1] Lucia Iglesias,[2] Manuel Bibes,[2] Wilfrid Prellier,[1] and Julien Varignon[1]

[1]Laboratoire CRISMAT, CNRS UMR 6508, ENSICAEN, Normandie Université,
6 boulevard Marechal Juin, F-14050 Caen Cedex 4, France

[2]Unite Mixte de Physique, CNRS, Thales, Université Paris Sud,
Université Paris-Saclay, F-91767 Palaiseau, France

**Abstract**

Nickel oxide superconductors offer an alternative playground for understanding the formation of Cooper pairs in correlated materials such as the famous cuprates. By studying the $La_{n+1}Ni_nO_{2n+2}$ phase diagram on the basis of hybrid and spin-polarized density functional theory simulations, we reveal the existence of charge and bond ordered (CBO) insulating phases that are quenched by doping effects, ultimately resulting in a metallic phase at the n=5 member. Nevertheless, the phonons associated with the CBO identified in the phase diagram remain sufficiently large to mediate Cooper pairs in $La_6Ni_5O_{12}$, yielding a computed critical temperature between $T_c$=11-19K consistent with the 13K observed experimentally in $Nd_6Ni_5O_{12}$. Thus, in order to identify the superconducting mechanism, extracting the relevant instabilities in the doping phase diagram of superconductors appears critical.

# I. Introduction

The absence of analogous compounds to cuprates has been detrimental to understanding the underlying mechanism behind high $T_c$ superconductivity. In that perspective, nickel-based superconductors were proposed decades ago [1,2] to host superconductivity (SC), but it was only realized in 2019 with the discovery of SC in infinite layered (IL) nickelates [3–8]. These materials possess the generic $R_{n+1}Ni_nO_{2n+2}$ chemical formula, also called the Reduced Ruddlesden-Popper (RRP) compounds, where R=La, Pr or Nd. They are based on n $RNiO_2$ building blocks stacked on top of each other and separated by a fluorite $RO_2$ spacing layer (see **Figure 1.a**). In these systems, the formal oxidation state (FOS) x of the Ni cations depends on the number of layers n present in the structure with x=(n+1)/n and hence varies between x=2 (n=1) and x=1 (n=∞) as discussed in Ref. [9].

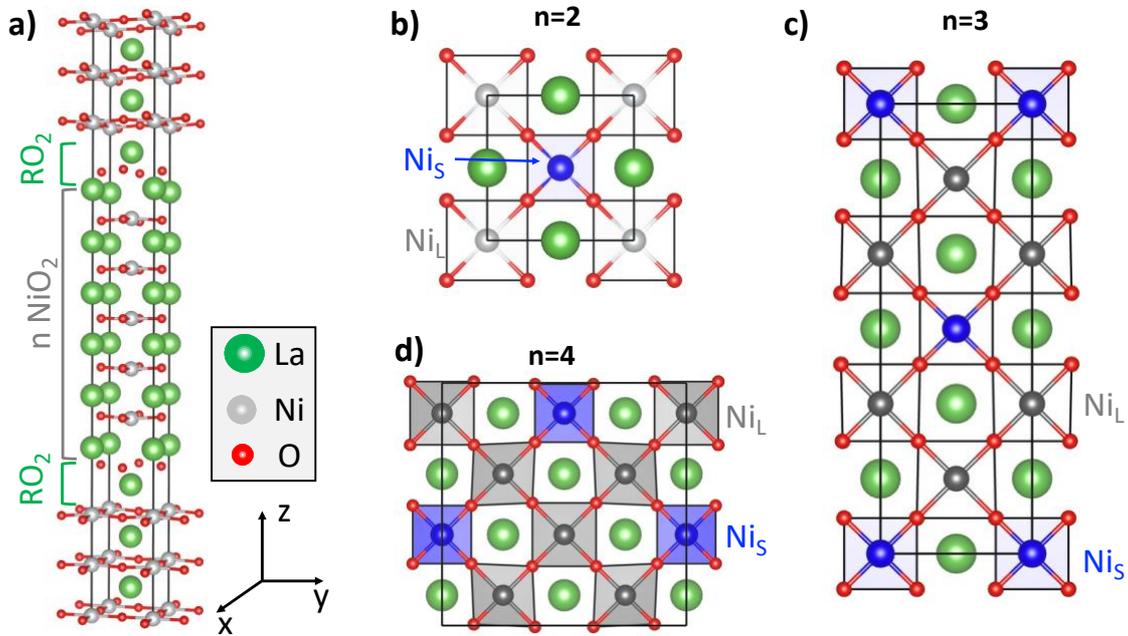

**Figure 1:** *Sketches of the high symmetry cell and related structural distortions exhibited by $R_{n+1}Ni_nO_{2n+2}$ compounds. **a)** High symmetry cell of reduced Ruddlesden-Popper compounds based on n $NiO_2$ planes stacked along the c axis and separated by $RO_2$ layers. **b-d)** Disproportionation distortions exhibited by n=2 **(b)**, n=3 **(c)** and n=4 **(d)** reduced Ruddlesden-Popper compounds.*

To date, most studies focus on the IL phase [10–18] and few of them discuss the RRP series, where either they focus on specific cases of the series [19–32] or they study the whole series albeit neglecting the spin and/or the structural degrees of freedom, leading to predict either all compounds to be metallic [33,34] or insulating if a long-range antiferromagnetic (AFM) order is imposed [35]. This behavior is at odds with experiments where the n=2 (x=1.5) compounds are insulating [25,27,30], the n=3 (x=1.33) case is found to be insulating at low temperatures [19,21–23], but a small pressure effect can turn it to a metal [20,24], and the n=5 (x=1.2) case is found superconducting [6]. This collection of experimental results shows that there has to be a transition between insulating to metallic character with increasing n in the RRP series of compounds. This is reminiscent of results of Ref. [9] where the transition to SC appears at x=1.2 for the doped infinite layered phase.

Nickelates are usually classified as non-conventional superconductors and several theoretical studies have indeed ruled out the existence of an electron-phonon coupling (EPC) for explaining the Cooper pairs formation [36,37]. However, these studies are performed with dramatic shortcuts such as (i) using highly symmetric primitive cells that prevent structural lowering events potentially coupled to electronic features, and/or (ii) involving an inappropriate description of exchange and correlation (*xc*) phenomena in Density Functional Theory calculations with using Local Density Approximation (LDA) or Generalized Gradient Approximation (GGA) and/or (iii) forcing similar occupancies between up and down electrons locally on each site (nonmagnetic solution, NM) despite the obvious presence of open shells on nickel atoms – experiments suggesting the presence of spins in the SC region [38]. All these shortcuts are nowadays well known to be irrelevant to model properties of transition metal oxides [39–42], including the trend in doping effects and emergence of superconductivity [9]. Furthermore, several experimental data analyses suggest an electron-phonon mechanism in these nickelates [43–45]. It is then natural to wonder "*what is the trend in electronic structure upon doping the RRP nickelates if one performs DFT calculations with including all relevant degrees of freedom*", "*is an electron-phonon coupling really ruled out at this DFT level*" and "*can we identify a common mechanism with other (oxide) superconductors?*"

In this article, we present a systematic study of the phase diagram of the $La_{n+1}Ni_nO_{2n+2}$ family of compounds by using Density Functional Theory (DFT) calculations that involve (i) local Ni spin formation; (ii) all relevant structural degrees of freedom and (iii) one of the highest levels of DFT functional for describing *xc* phenomena before explicitly including dynamical correlation effects. We reveal the existence of multiple charge and bond orderings (CBO) as well as insulating states in the phase diagram which disappears at the n=5 case, finding a metallic compound. We thus reproduce the experimental crossover from insulating (n=1, 2, 3, and 4) to metallic (n=5). Within the metallic compound, we demonstrate that the bond disproportionation vibrations associated with the CBO identified for n=2 and n=3 compounds are sufficient to reproduce an electron-phonon coupling constant λ=0.61, yielding a critical temperature $T_c$ estimated between 19 and 11 K. This is in very good agreement with the experimental $T_c$ of 13 K measured in $Nd_6Ni_5O_{12}$ [6], further supporting the phonon-mediated mechanism identified in infinitely layered nickelates [9] and proposed on the basis of experimental data analyses [43,44]. These results highlight that the phase diagram of nickelates and their SC properties must be understood starting from the n=1 member and not the n=∞, allowing to identify the existence of a CBO phase as a common root with bismuthate and antimonate superconductors [46–50].

## II. Method

*The choice of the exchange-correlation functional:* in order to sufficiently amend self-interaction errors (SIE) inherent to the practical implementation of Density Functional Theory (DFT), the choice of the *xc* functional is a critical step, especially for modelling transition metal oxides (TMOs). It is now well established that local and semi-local *xc* functionals such as LDA and GGA (*i.e.* rung 1 and 2 on the Jacob's ladder [51]) cannot capture the trends in structural distortions and metallic or insulating character of TMOs, and hence it cannot be employed to model their doping effects and superconducting properties [39–42]. One can employ meta-GGA *xc* functionals (rung 3, semi-local functional) such as the Strongly Constrained and Appropriately Normalized (SCAN) [52] or DFT+U/hybrid methods (rung 4) that are non-local operators of the density matrix. The DFT+U method has a disadvantage for doped compounds in which cations can exhibit multiple formal oxidation states (FOS): the U parameter, preventing the delocalization of electrons, is FOS dependent. The meta-GGA *xc* functionals offer a parameter free technic but it can fail at capturing (or strongly underestimate) the bands splitting and band gap of compounds, especially for compounds with low FOS [9,39,53]. In this study, we have employed the hybrid Heyd-Scuseria-Ernzerhof HSE06 functional [54] which is one of the highest level of DFT before explicitly treating dynamical correlation effects. It has two advantages: (i) calculations remain at a parameter free level and (ii) it offers a correct estimation of bands splitting that is a critical aspect to evaluate the electron-phonon coupling as we discuss below. We have used the standard HSE06 parametrization with 25% of exact Hartree-Fock exchange and a range separation of 0.2 Å. We emphasize that the HSE06 functional involves a Perdew-Burke-Ernzerhof (PBE) correlation term (GGA functional) [55], and hence it does not improve that description. Instead, it corrects the exchange part with exact Hartree-Fock exchange for amending the spurious SIE term. Although the HSE06 can sometimes over localize electrons and exaggerate the insulating state such as in cuprates [56,57], the HSE06 correctly captures the insulating state of the n=1 RRP nickelate member (as shown later in the manuscript) as well as the metallic nature of the n=∞ member [53]. Hence one expects that the trend in structural distortions and insulating or metallic character of the intermediate RRP members is correctly captured.

***The critical role of local spin formation:*** as discussed in the introduction, literature abounds of theoretical studies ruling out the existence of an EPC in nickelate superconductors [36,37]. However, these studies are performed with non-spin polarized calculations (NM), preventing local spin formation by construction. This is rather counter intuitive, notably for Ni cations adopting 1+ ($3d^9$) to 2+ ($3d^8$) formal oxidation state (FOS) for which there is obviously unpaired electrons. Such a shortcut was previously shown to be irrelevant to capture the physical properties of TMOs [39–42], including the superconducting mechanism in infinite layered nickelates [9]. As discussed in Ref. [9], the NM approximation for doped infinite layered nickelates produces solutions at least 1800 k above any spin-polarized solutions. Even though one involves the highest level of description of correlation effects with Green's function, it remains an improvement of an erroneous/irrelevant solution [37,58].

To circumvent this shortcut, we have employed a ferromagnetic (FM) order since it is the lowest spin-order state allowing coupled electronic-structural instabilities to be captured, such as disproportionation and Jahn-Teller effects [39]. It also has the advantage to let us access the sole role of doping effects on the trend in structural distortions and insulating/metallic transitions without introducing antiferromagnetic (AFM) correlations that can make bands more compact and force electronic instabilities toward disproportionation and/or Jahn-Teller effects [59]. However, we have performed additional checks using various AFM orders and the overall results are unchanged.

***The choice of the unit cells:*** Several studies restrict calculations to highly symmetric primitive unit cells preventing structural lowering events to occur. Again, these shortcuts cannot account for potential structural distortions altering the band structure and bands splitting [42]. We have therefore employed sufficiently large supercells allowing potential localization of added electrons (*i.e.* a polaronic states). At this stage, this is critical to use cells with integer number of electrons to allow such phenomena, instead of using a simple virtual crystal approximation (VCA) with a fractional number of electrons per primitive cell –hence preventing polaronic states to form by construction. Our supercells correspond to ($\sqrt{2}$, $\sqrt{2}$, 1) (n=1), ($\sqrt{2}$, $\sqrt{2}$, 1) (n=2), (3, 1, 1) (n=3) and ($2\sqrt{2}$, $2\sqrt{2}$, 1) (n=4) and (5, 1, 1) (n=5) cells with respect to the high symmetric I4/mmm cell of RRP compounds.

***Estimation of the electron-phonon coupling and critical temperature:*** to date, there is no practical implementation of the calculation of the electron-phonon coupling with local spin degree of freedom [60]. Thus, one has to circumvent the problem and compute manually the electron-phonon coupling. To that end, we have considered the phonons that are intrinsically strongly coupled to electronic features such as bond disproportionation [9,39,47–49]. The electron-phonon coupling constant $\lambda$ can be evaluated as the sum of the mode $\nu$ and momentum $\vec{q}$ dependent coupling constant $\lambda_{\vec{q},\nu}$ [61,62]:

$$\lambda = \frac{1}{N_{\vec{q}}} \sum_{\vec{q},\nu} \lambda_{\vec{q},\nu} \quad \textbf{(equation 1)}$$

where $N_{\vec{q}}$ is the number of $q$ points. Additionally, $\lambda_{\vec{q},\nu}$ in the double delta approximation is given by [63–66]:

$$\lambda_{q,\nu} = \frac{2}{N(E_F)} \sum_{\vec{k}} \frac{1}{\omega_{\vec{q},\nu}} \left| M^{\nu}_{\vec{k},\vec{k}+\vec{q}} \right|^2 \delta(E_{\vec{k}} - E_F) \delta(E_{\vec{k}+\vec{q}} - E_F) \quad \textbf{(equation 2)}$$

where $N(E_F)$ is the density of states at the Fermi level $\varepsilon_F$, the electron energies $E_{\vec{k}}$ and $E_{\vec{k}+\vec{q}}$ with momentum $\vec{k}$ and $\vec{k}+\vec{q}$, respectively, $\omega_{\vec{q},\nu}$ is the phonon mode frequency at a given momentum $\vec{q}$ and $M^{\nu}_{\vec{k},\vec{k}+\vec{q}}$, is the electron phonon matrix element which can be defined as :

$$M^{\nu}_{\vec{k},\vec{k}+\vec{q}} = \sum_j \left( \frac{\hbar^2}{2 M_j \omega_{\vec{q},\nu}} \right)^{1/2} e^{\nu}_{\vec{q},j} \left\langle \vec{k}+\vec{q} \left| \frac{\delta V}{\delta u^{\nu}_{\vec{q},j}} \right| \vec{k} \right\rangle \quad \textbf{(equation 3)}$$

with $M_j$ the mass of the moving atom, $e^{\nu}_{\vec{q},j}$ is the j$^{\text{th}}$ component of the polarization vector and $\left\langle \vec{k}+\vec{q} \left| \frac{\delta V}{\delta u^{\nu}_{\vec{q},j}} \right| \vec{k} \right\rangle$ is the reduced electron phonon matrix element (REPME). Since we are interested only on the electronic states close to the Fermi level, we can change the delta functions in **equation 2** by the density of states at the Fermi level $N(E_F)$ obtaining :

$$\lambda_{\vec{q},\nu} = 2 N(E_F) \sum_{\vec{k}} \frac{1}{\omega_{\vec{q},\nu}} \left| M^{\nu}_{\vec{k},\vec{k}+\vec{q}} \right|^2 \quad \textbf{(equation 4)}$$

Introducing eq. 4 back into **equation 1**, we obtain that the electron phonon coupling constant can be expressed as :

$$\lambda = \frac{2 N(E_F)}{N_{\vec{q}}} \sum_{\vec{k},\vec{q},\nu} \frac{1}{\omega_{\vec{q},\nu}} \left| M^{\nu}_{\vec{k},\vec{k}+\vec{q}} \right|^2 \quad \textbf{(equation 5)}$$

For modes that involve only a single type of atomic species, such as the bond disproportionation modes that involve only oxygen anions with mass $M_O$, the electron phonon matrix element in **equation 3** can be simplified into :

$$M^{\nu}_{\vec{k},\vec{k}+\vec{q}} = \left( \frac{\hbar^2}{2 M_O \omega_{\vec{q},\nu}} \right)^{1/2} e^{\nu}_{\vec{q},O} \left\langle \vec{k}+\vec{q} \left| \frac{\delta V}{\delta u^{\nu}_{\vec{q},O}} \right| \vec{k} \right\rangle \quad \textbf{(equation 6)}$$

Regarding the REPME, one can evaluate its magnitude for a given phonon branch and $\vec{q}$ vector by finite displacements and measuring the difference in the Kohn-Sham eigen energies [61–63,67] :

$$\left\langle \vec{k}+\vec{q} \left| \frac{\delta V}{\delta u^{\nu}_{\vec{q},O}} \right| \vec{k} \right\rangle = \frac{E_{\vec{k}+\vec{q}} - E_{\vec{k}}}{\Delta u^{\nu}_{\vec{q},O}} \quad \textbf{(equation 7)}$$

with $\Delta u^{\nu}_{\vec{q},O}$ being a finite amplitude of the given phonon mode frozen in the material. In the case of a mode involving a single atomic species, one can write the following identity:

$$e_{\vec{q},O}^{\nu}\left\langle\vec{k}+\vec{q}\left|\frac{\delta V}{\delta u_{\vec{q},O}^{\nu}}\right|\vec{k}\right\rangle = \frac{E_{\vec{k}+\vec{q}}-E_{\vec{k}}}{|Q_{\vec{q},\nu}|} \quad \textbf{(equation 8)}$$

where $Q_{\vec{q},\nu}$ is the phonon mode amplitude. Now since we are interested mainly about physics around the Fermi level, we can evaluate the change in the eigen states as the band splitting $\Delta E_b = \left(\frac{E_{\vec{k}+\vec{q}}-E_{\vec{k}}}{2}\right)$ and the REPME $D_{\vec{q}}^{\nu} = \frac{\Delta E_b}{2|Q_{\vec{q},\nu}|}$. Assuming an isotropic coupling (*i.e.* bands splitting are homogeneous through the entire Brillouin zone), we can drop the $\vec{k}$ dependence on the sum in **equation 5**, and we would obtain

$$\lambda = \frac{2N(E_F)}{N_{\vec{q}}}\sum_{\vec{q},\nu}\frac{\hbar^2}{2M_O\omega_{\vec{q},\nu}^2}\left|D_{\vec{q}}^{\nu}\right|^2 \quad \textbf{(equation 9)}$$

In practice, the DFT calculations are used to extract (i) the bands splitting $\Delta E_b$ associated with the introduction of a finite $(\vec{q},\nu)$ phonon mode and (ii) the frequency $\omega_{\vec{q},\nu}$ of the mode. Point (i) is highly critical as one needs to accurately estimate the bands splitting. It is well known that local and semi-local *xc* functionals underestimates band gaps – and hence bands splitting – of materials, including those of highly uncorrelated semi-conductors such as Si and Ge. Hence one has to use a DFT functional accurately capturing this quantity. At this stage, hybrid DFT represents the best compromise between accuracy and a parameter free technique. The evaluation of point (ii) is rather straightforward: by fitting the total energy E versus mode amplitude $Q_{\vec{q},\nu}$ of a $(\vec{q},\nu)$ phonon starting from the ground state structure with a polynomial expression of the form $E(Q_{\vec{q},\nu}) = aQ_{\vec{q},\nu}^2 + bQ_{\vec{q},\nu}^4$, one directly extracts the mode frequency $\omega_{\vec{q},\nu}$ as

$$\omega_{\vec{q},\nu} = \sqrt{2a/M_O} \quad \textbf{(equation 10)}$$

for $(\vec{q},\nu)$ modes involving only O motions – recalling that the energy of an harmonic oscillator is given by $\frac{1}{2}M_O\omega_{\vec{q},\nu}^2 Q_{\vec{q},\nu}^2$.

Finally, one estimates the critical temperature $T_c$ of the material by using the Mc Millan-Allen equation

$$T_c = \frac{\hbar\omega_{log}}{1.2}\exp\left(-\frac{1.04\lambda}{\lambda-\mu^*(1+0.62\lambda)}\right) \quad \textbf{(equation 11)}$$

where µ* is the screened Coulomb potential with typical values between 0.1 and 0.15 and $\omega_{log}$ is the logarithmic frequency which can be estimated for a finite number of frequencies as

$$\omega_{log} = \left(\prod_i^n \omega_i\right)^{1/n} \quad \textbf{(equation 12)}$$

***Other details of the method:*** First-principles DFT simulations are performed with the CRYSTAL17 software package [68]. Structural optimizations (lattice parameters plus atomic positions) are performed until the root mean square of forces and estimated displacements are lower than $3 \cdot 10^{-4}$ Ha.Bohr$^{-1}$ and 0.0012 Bohr, respectively. The convergence on the total energy is set to $10^{-9}$ Hartree. Gaussian basis sets with core electrons treated in pseudopotentials are used for Ni and La cations, extracted from Refs. [69–71], and an all-electron basis set is used for O anions and taken from Ref. [72]. Brillouin zone are sampled with 8x8x8 (n=1), 4x4x8 (n=2), 3x3x3 (n=3), 4x4x2 (n=4) and 4x4x4 (n=5) points. All calculations are performed within the primitive cell as implemented in CRYSTAL17. The mesh was increased to 32x32x32 points for density of states calculations for the n=5 member. Symmetry mode analysis is performed with the AMPLIMODES application [73,74] implemented in the Bilbao Crystallographic Server and ISODISTORT from the ISOTROPY software packages [75,76].

### III. Results and discussion

#### 1. The failure of the NM approximation

We first evidence the need to include the spin degree of freedom by performing a study of the phase diagram (n=1-5), using the non-spin-polarized (NM) approximation. Results of structural relaxations are provided in **Table 1**. We find that (i) all compounds present conducting behavior except the first member (n=1) that is insulating and (ii) all ground states are achieved with a high symmetry I4/mmm crystal structure as found in previous studies [33,34]. This in contrast to experimental data available for the n=1 compound and with spin-polarized DFT performed here [77–79] showing an Aema space group with octahedral rotations.

| n | FOS | phase | Space Group | a (Å) | b (Å) | c (Å) | $E_g$ (eV) | ΔE (meV/fu) | $\mu_{NiL}$ | $\mu_{NiS}$ | $Q_{Boc}$ (Å/f.u) |
|---|---|---|---|---|---|---|---|---|---|---|---|
| 1 | 2+ | NM | I4/mmm | 3.70 | 3.70 | 13.45 | 1.36 | 0 | - | - | 0 |
| | | FM | Aema | 5.57 | 5.46 | 12.52 | 2.80 | -28 | 1.73 | 1.73 | 0 |
| | | exp. | Aema [78] | 5.53 | 5.43 | 12.49 | 1.51 [80] - 4 [81] | - | 1.48 [77,82] | - | 0 |
| 2 | 1.5+ | NM | I4/mmm | 3.95 | 3.95 | 18.93 | 0 | 0 | - | - | 0 |
| | | FM | Ammm | 5.62 | 5.62 | 18.88 | 1.24 | -263 | 0.93 | 0.09 | 0.09 |
| | | exp. | I4/mmm [25] | - | - | - | Ins. [30,83] | - | - | - | >0 [25] |
| 3 | 1.33+ | NM | I4/mmm | 3.96 | 3.96 | 25.36 | 0 | 0 | - | - | 0 |
| | | FM | Fmmm | 16.88 | 5.63 | 23.31 | 1.00 | -313 | 0.93 | 0.15 | 0.07 |
| | | exp. | F4/mmm [84] | 5.61 | 5.61 | 26.09 | Ins. [22,24] | - | >0 [23,85] | >0 | >0 [22,23] |
| 4 | 1.25+ | NM | I4/mmm | 3.96 | 3.96 | 31.84 | 0 | 0 | - | - | 0 |
| | | FM | Cmcm | 11.27 | 11.27 | 31.71 | 1.49 | -398 | 0.93 | 0.17 | 0.05 |
| | | exp. | - | - | - | - | - | - | - | - | - |
| 5 | 1+ | NM | I4/mmm | 3.96 | 3.96 | 38.33 | 0 | 0 | - | - | 0 |
| | | FM | I4/mmm | | | | 0 | 0 | 0.77 | 0.77 | 0 |
| | | exp. | I4/mmm | - | - | - | - | - | - | - | 0 |

*Table 1: Key properties of relaxed RRP compounds with a non-spin polarized (NM) and ferromagnetic (FM) solution for the n=1-5 members. Experimental data when available are provided. Results entail the formal oxidation state (FOS), the space group, lattice parameters (in Å), the band gap $E_g$ (eV), the energy difference between the FM and NM solutions (in meV/f.u), the magnetic moments of the different type of Ni cations (in $\mu_B$) and the amplitude (in Å/f.u) of disproportionation modes $Q_{Boc}$ extracted from a symmetry mode analysis with respect to a high symmetry I4/mmm cell.*

The n=1 ($Ni^{2+}$) compound is identified to be a band insulator with a band gap of 1.36 eV, originating from a $3d^8$ low spin (S=0) electronic configuration enforced by the NM approximation (see **Figure 1**). This means that the Hund's coupling is neglected and the crystal field splitting of doubly occupied $3d$ shell for $Ni^{2+}$ cation is responsible of a band gap opening (see **Figure 1.a**). However, this is in contrast with experiments that reveal the existence of a magnetic moment of 1.48 $\mu_B$ per Ni cation in $La_2NiO_4$ [77,82], thus suggesting a high spin configuration as depicted in **Figure 1.b**. The high symmetry cell identified with DFT is also at odds with the experimental structure exhibiting $a^-a^-c^+$ octahedral rotations with Aema space group [77–79]. Even though we start from this structure, it relaxes back to a I4/mmm cell with no distortion. Regarding higher members of the series, the observation of a metallic character for n =2 or n=3 disagrees with the experimental observation of an insulating state at low temperatures [22,24,30,83].

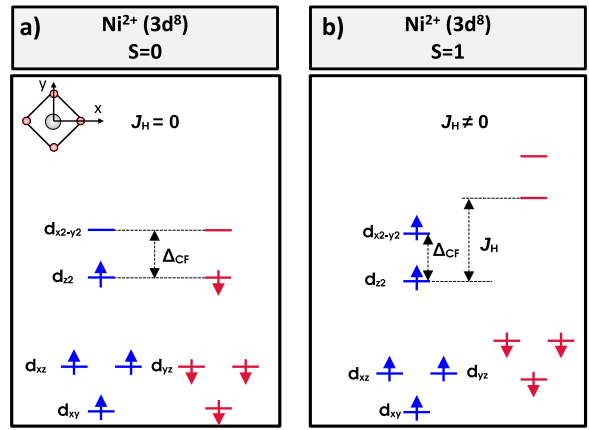

*Figure 2: Diagram of the crystal field splitting in on $Ni^{2+}$ cations with a $3d^8$ electronic configuration in the case of $La_2NiO_4$ for two different situations: **a**) Hund's coupling is neglected and a NM solution is employed in the DFT calculations producing a S=0 spin state; **b**) Hund's coupling is included and the spin degree of freedom is allowed in the DFT calculations producing a S=1 spin state.*

This general failure of the NM approximation is grounded on the fact that Hund's rule is broken and important terms in the energy are not well accounted. We evidence our claims by performing spin-polarized FM calculations in the high symmetry I4/mmm cell and compute the total energy difference between the two types of calculations. This allows us to reveal the sole effect of the spin degree of freedom in the energy of the system without including any lattice distortions that may appear in the global ground state. We report the energy difference $\Delta E_{NM-FM}$ between the NM and FM solutions as a function of n in **Figure 2**. A massive energy gain $\Delta E_{NM-FM}$ of at least 375 meV/NiO$_2$ motif for all n=1-5 RRP compounds, signaling that a huge part of the energy of the system is not well accounted for if one ignores the local spin formation. Thus, we conclude here that (i) local Ni spin formation in these nickelates is a key factor that should not be overlooked in the DFT simulations and (ii) even though we treat exchange correlation phenomena at one of the highest level in DFT simulations with involving hybrid *xc* functionals, the NM solutions remain irrelevant for modeling the electronic properties of oxide superconductors.

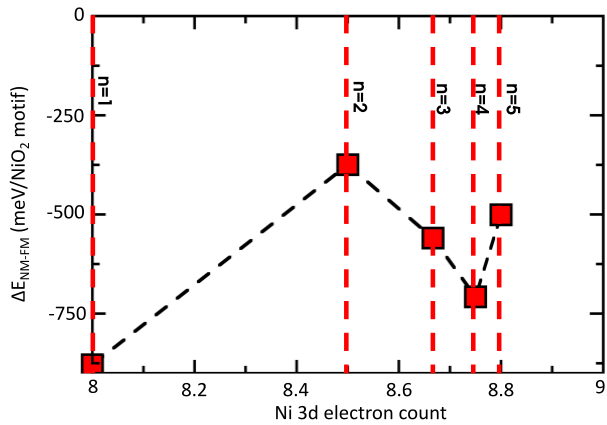

*Figure 3*: Energy difference $\Delta E_{NM-FM}$ between a non-spin- polarized (NM) and ferromagnetic (FM) order (in meV/NiO$_2$ motif) as a function of the formal Ni 3d occupancy in the reduced Ruddlesden-Popper compound. Red dashed lines indicate the position of each member n of the RRP series.

### 2. The ground state properties with a spin-polarized solution

Key results of structural relaxations for n=1 to n=5 members with a FM order are reported in **Table 1**. For all members, we identify structural lowering events producing large energy gains up to 400 meV/f.u. at the exception of the n=5 member that keeps its high symmetry I4/mmm cell.

The n=1 is found to be an insulator with a gap located between majority and minority spin channels of Ni cations associated with a 3d$^8$ (Ni$^{2+}$) high spin electronic configuration (see Figure 4.a). Unlike the NM solution, the spin-polarized solution can stabilize the Aema space group that is observed experimentally [77–79]. The computed magnetic moments as well band gap values and crystallographic cell are in good agreement with experiments ($\mu_{Ni}$=1.48 $\mu_B$ [77,79] and E$_g$=1.51 - 4 eV [80,81]), confirming the quality and reliability of the HSE06 functional for the present studies of doping effects.

For the n=2 to n=4, we observe large energy gains related to the appearance of sizable bond disproportionation distortions amplitude $Q_{B_{oc}}$ producing different local environments for Ni cations (see **Figures 1.b** to **1.d**). It produces two types of Ni cations sitting in a compressed and extended $O_4$ complex, and labeled $Ni_S$ and $Ni_L$, respectively, with a ratio between large and small complexes able to accommodate the Ni mixed valence states. The resulting effect of the disproportionation effect is to produce different magnetic moments with $Ni_S$ approaching 0 $\mu_B$ and $Ni_L$ bearing a magnetic moment of roughly 1 $\mu_B$. It points toward Ni cations adopting a $3d^8$ low spin ($Ni_S^{2+}$) and $3d^9$ ($Ni_L^+$) electronic configurations, respectively. The resulting effect of the $B_{oc}$ modes is to open a band gap in the materials with band edges formed between $Ni_L^+ d$ and $Ni_S^{2+}$ - $d$ states in the n=2, 3 and 4 RRP compounds (see **Figures 4.b-d**). These results are compatible with experiments that show an insulating character for n=2 and n=3 members – the n=4 compound has not been synthesized to the best of our knowledge. Furthermore, the existence of charge orderings between $Ni^+$ and $Ni^{2+}$ for n=2 and n=3 RRP is confirmed experimentally [19–22,25,30]. Their existence was also predicted theoretically for n=2-3 [31,86,87]. The propensity of Ni cations to undergo disproportionation effects in infinite layered phase was also proposed on the basis of more elaborated Dynamical Mean Field Theory (DMFT) calculations [14,88], confirming the reliability of the DFT results albeit not including dynamical correlation effects. It was also suggested experimentally in O rich doped n=∞ members [89]. Finally, the existence of disproportionation effects of the 1.5+ formal oxidation state of Ni cations in $La_3Ni_2O_6$ is in sharp agreement with our previous simulations on half-doped infinite layered nickelates [9]. However, we should point out that we could not find an experimental structure able to adopt a charge ordering – a primitive I4/mmm cell reported for the n=2 and n=3 situations is unable to account for the $Ni^+/Ni^{2+}$ charge and bond ordered character and *de facto* its insulating state [25]. The I4/mmm space group has also been precisely challenged for the n=3 situation by X-ray diffraction that reveal a stripe charge order of 2 $Ni^+$ and 1 $Ni^{2+}$ in (ab) planes stacked along the c axis – as we do observe on our simulations [22]— hence incompatible with a I4/mmm primitive cell symmetry.

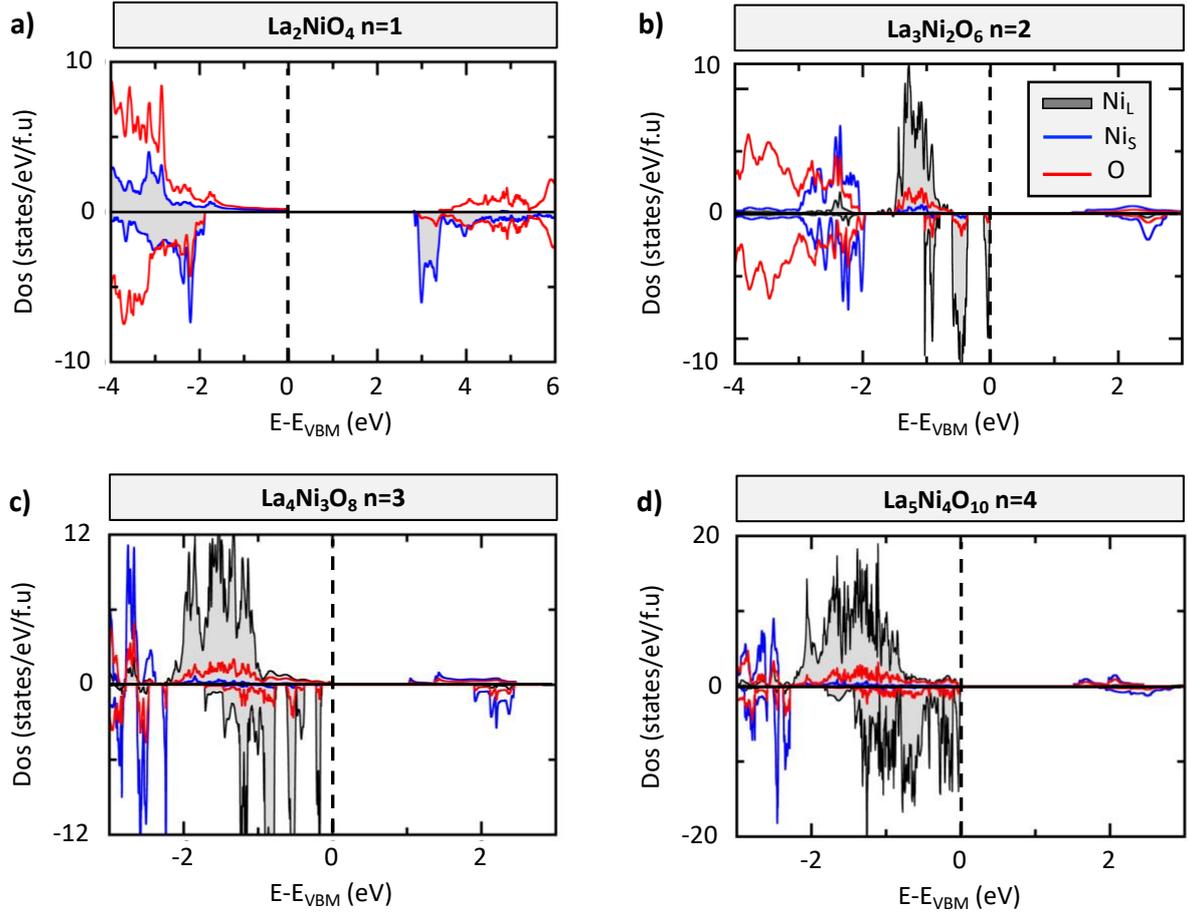

*Figure 4:* Electronic properties of the reduced Ruddlesden- Popper material from n=1 to n=4. Projected identity of states on $Ni_L$ (filled grey area), $Ni_S$ (blue line), and O (red line) states for the n=1 (*a*), n=2 (*b*), n=3 (*c*) and n=4 (*d*) RRP members. The vertical dashed line represents the valence band maximum (VBM).

The n=5 member is found to be metallic with an I4/mmm structure. Even though we start from a charge ordered cell, it spontaneously vanishes and relaxes back to a high symmetry metallic state. This observation of a metallic state for n=5 is in agreement with the insulating to metallic crossover between n=1-3 and n=5 members observed experimentally. We conclude here that n=2-4 RRP compounds are prone to exhibit CBO, resulting in an insulating phase.

### 3. Origin of the various bond disproportionation modes

*Potential energy surfaces for the FM order:* In order to understand the origin of the charge orderings observed in the different nickelate members, we plot the potential energy surface (PES) associated with each type of disproportionation mode starting from the highly symmetric undistorted I4/mmm cell with identical $O_4$ complexes for each member of RRP n = 2 to n = 4. Results are displayed in **Figure 5**. We start by analyzing the PES for the FM state. The n=2 type of distortion shows a shifted single well potential to non-zero values of the mode and produces a large energy gain of -117 meV/$NiO_2$ motif. It signals the presence of an electronic instability associated with the 1.5+ formal oxidation state that prefers to transform to more stable 1+ and 2+ in the ground state. The potentials for n=3 and n=4 are slightly different with respect to the n=2 mode by showing a double well potential shape and lower energy gains, reaching at most -40 meV/$NiO_2$ motif. The existence of a double well potential and lower energy gains suggest a partial screening of the electronic instability. Such screening originates from doping effects that alter the band compacity of the Ni $d$ states, thereby weakening the electronic instability as explained in Ref.[54]. We indeed observe a global increase of the bandwidth W from W =1.5 eV (n=2) to W =2.2 eV for n=3 and n=4 in the starting I4/mmm undistorted cell.

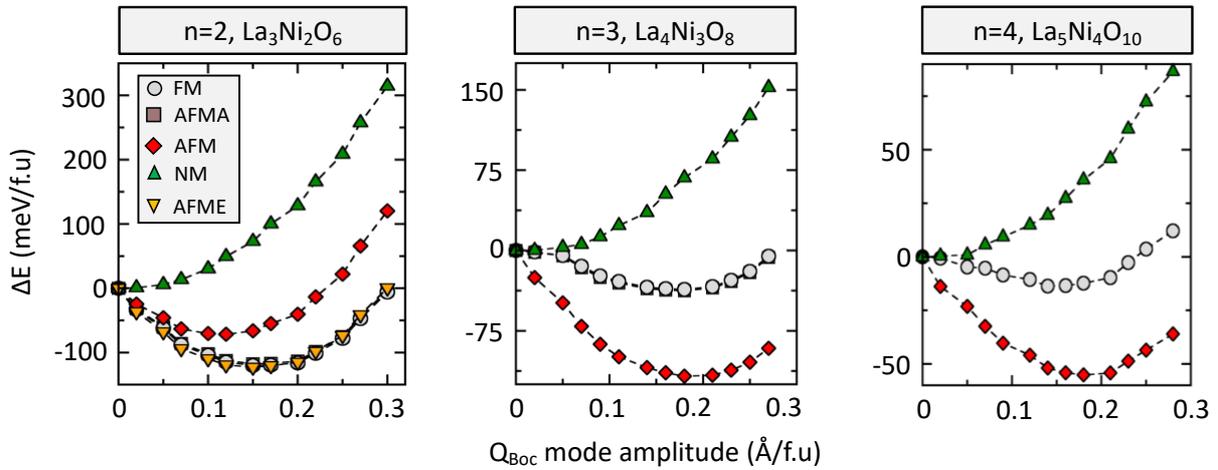

*Figure 5: Potential energy surface associated with the various bond disproportionation modes. Evolution of the total energy ΔE (in meV/f.u) as a function of the $B_{oc}$ mode amplitude $Q_{Boc}$ (in Å/f.u) in the n=2 (left panel), n=3 (middle panel) and n=4 (right panel) RRP compounds. The reference energy is taken at $Q_{Boc}$=0. Calculations are performed for a ferromagnetic interactions (FM, grey filled circles), (ab) plane FM interactions with antiferromagnetic interactions between consecutive planes along the c axis (AFMA, brown filled squares), pure AFM interaction between all nearest neighbors in all directions (AFM, red filled diamonds), AFM interaction between second nearest neighbor in the (ab)-plane with FM couplings along c (AFME, orange filled down triangles) and non-spin polarized solutions (NM, green filled triangles).*

***Results are independent of magnetic orders:*** By using AFM orders (**Figure 5**), the extra correlations can improve the band compacity of Ni $d$ states with respect to an FM order [59]. The resulting effect is to shift the single well potential to lower energies, signaling a stronger tendency to develop disproportionation effects for n=2, n=3 and n=4 potential energy surfaces.

***Local spin formation is a critical ingredient:*** We performed additional calculations of the PES using the NM approximation. Unlike any of tested spin polarized solutions, the PES for the n=2, 3 and 4 compounds exhibit single well potentials centered at 0, indicating that the materials are not willing to undergo any disproportionation effects. This is in contrast with experiments and other theoretical calculations that show the stabilization of $Ni^+$ and $Ni^{2+}$ in the compounds (see references therein). This observation is reminiscent of the $RNiO_3$ situation where non spin-polarized DFT calculations could not predict the stabilization of the disproportionation effects, at odds with experiments and spin-polarized DFT calculations [39].

***Check of the existence of an electronic instability:*** the existence of an electronic instability can be confirmed by performing two sets of calculations in the high symmetry cell for the n=2 compound without any lattice distortion and with a FM order in which (i) we initially force an equal occupancy of all neighboring Ni cations and (ii) a calculation where we initially force neighboring Ni cations to have either a $d^8$ or a $d^9$ configuration. After the variational self-consistency, we observe that (ii) produces an energy gain of 35 meV/$NiO_2$ motif, confirming the existence of the electronic instability of Ni $3d^{8.5}$ configuration. The material then develops a lattice distortion to accommodate the charge ordering with a ratio 1:1 of $Ni^+$:$Ni^{2+}$ and produces an insulating state.

***The disproportionation effects are independent of the xc functional:*** although one may argue that the HSE06 functional may over localize electrons and enforce insulating states instead of metals, such as in doped cuprates, we have checked that the tendency toward disproportionation effects is independent of the DFT *xc* functional. Ref. [9] already identified the tendency toward disproportionation effect of $La_3Ni_2O_6$ using the meta-GGA SCAN functional. It signals the presence of a strong electronic instability toward disproportionation even though one implies a "lower level" DFT xc functional on the Jacob's ladder. We conclude here that these nickelates are governed by a large electron-phonon coupling able to produce a charge ordering, electron localization, and open a band gap. This is observed only if relevant degrees of freedom such as the local Ni spin formation are included in the simulation.

### 4. *The metallic state is at the vicinity of a charge ordered phase*

The weakening of the tendency toward disproportionation effects of Ni cations upon doping the compounds can also be tracked by quantifying the amplitude associated with the relevant disproportionation $B_{oc}$ modes in the ground state structures (**Figure 6**). We observe that upon increasing the number of layers n, from n=2 to n=5, there is a clear weakening of the $B_{oc}$ modes amplitude until it vanishes at n=5. Therefore, all types of disproportionation effects are quenched by electron doping the nickelates. Thus, the metallic region and observed superconducting RRP compound (*i.e.* n=5) sit in the vicinity of a charge and bond ordered phase. This situation is reminiscent of antimonate (bismuthate) superconductors $Ba_{1-x}K_xSbO_3$ ($Ba_{1-x}K_xBiO_3$) in which hole doping (i) suppresses disproportionation instabilities associated with $Sb^{4+}$ ($Bi^{4+}$) cations at x=0 and (ii) enables a superconducting regime once the material is sufficiently doped [46,48,50]. Although one may think that the insulator to metal transition could be related to dimensionality crossover as in iridates $SrIrO_3$ (metal) and $Sr_2IrO_4$ (insulator) [90,91] – going from n=2 to n=∞ members allows to tune the distance between Ir and apices O and therefore alter the crystal field splitting of Ir $d$ states –, the nickelates systems are different since there are no apices oxygen. Hence the crystal field is not sizably altered by increasing the number of layers in the RRP series. We thus conclude here that the electronic instability and its subsequent lattice distortion in nickelates are sufficiently weakened by electron doping in order to prevent electron localization, consequently resulting in a metallic state.

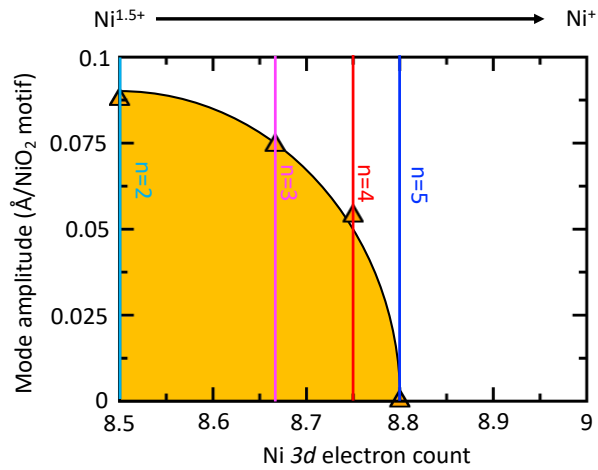

*Figure 6:* Amplitude of the breathing $B_{oc}$ mode (in Å/f.u) extracted from a symmetry mode analysis as a function of the Ni 3d electron count.

### 5. Superconductivity is mediated by an electron-phonon coupling

Although the electron-phonon coupling (EPC) is sufficiently weakened by doping effects so that it cannot produce electron localization and produce an insulating state, the disproportionation modes can still be highly coupled to the electronic structure of the metallic n=5 case. This enables the possibility of a phonon-mediated superconductivity in this metallic compound. We then evaluate the EPC of the $B_{oc}$ modes using **equation 9**. This is performed by freezing the n=2 and n=3 disproportionation modes in the n=5 structure – we could not freeze the n=4 disproportionation mode in the n=5 case due to the required supercell size that becomes too large.

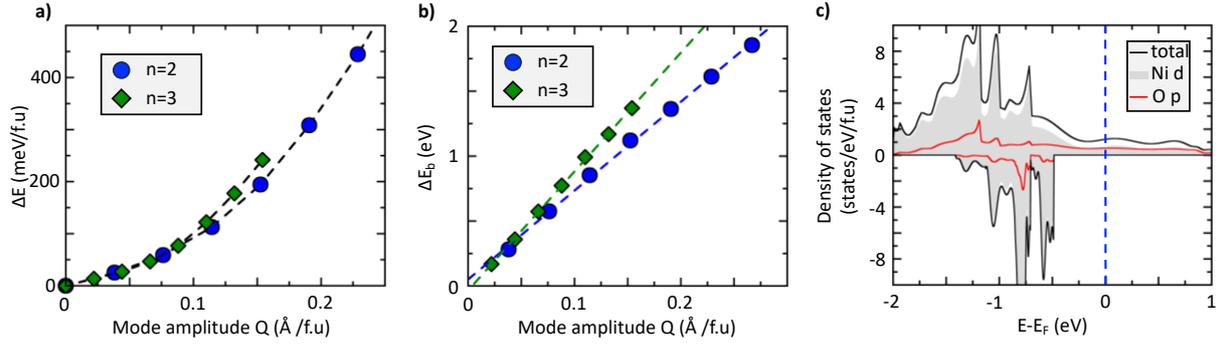

**Figure 7**: *Superconducting quantities in the metallic state of $La_6Ni_5O_{12}$.* **a)** *Energy difference ΔE (in meV/f.u) upon freezing amplitude (in Å) of the n=2 (blue filled circles) and n=3 (green filled diamonds) $B_{oc}$ modes in the ground state structure. The zero energy is set the ground state energy.* **b)** *Bands splitting $ΔE_b$ (in eV) induced by freezing amplitude (in Å/f.u) of the n=2 (blue filled circles) and n=3 (green filled diamonds) $B_{oc}$ modes in the ground state structure.* **c)** *Density of states (in states/eV/f.u) associated with the ground state structure. A coarse kmesh made of 32x32x32 points is used to sample the Brillouin zone.*

From the PES associated with the n=2 and 3 modes starting from the relaxed n=5 ground state (**Figure 7.a**), we can fit the curves and use **equation 10** to extract $ω_2$=69 meV and $ω_3$=66 meV for the modes with the $\vec{q}$ vector of the n=2 and n=3 $B_{oc}$ modes, respectively (*i.e.* (1/2,1/2,0) and (1/3,1/3,0)). The n=2 mode is compatible with the frequency computed for half-doped infinite layered nickelates in Ref. [9] to $ω_2$=65 meV. This frequency is also very similar to similar disproportionation modes appearing in bismuthates and antimonates [47–49,92]. Regarding the evaluation of the reduced electron-phonon matrix element (REPME) $D_{\vec{q},υ}$, we have frozen finite amplitudes Q of n=2 or n=3 phonon modes in the n=5 ground state structure. We observe a splitting of bands dispersing around the Fermi level (**Figure 8**). The splitting is homogeneous along the $Γ - X - M - Γ - Z$ path, meaning that the effect of modes on the band dispersion is isotropic. We further check that the bands splitting $ΔE_b$ are linear with the amplitude of the displacements (**Figure 7.b**) – thereby one is still in the harmonic regime –, and from a linear fit of the curves, we extract that $D_2$=7.8 eV/Å and $D_3$=8.6 eV/Å for the n=2 and n=3 modes, respectively. Finally, we extract the ground state density of states at the Fermi level $N(E_F)$ of the n=5 material by sampling the Brillouin zone with a kmesh made of 32x32x32 points and extract $N(E_F)$=0.607 states/eV/f.u/spin channel.

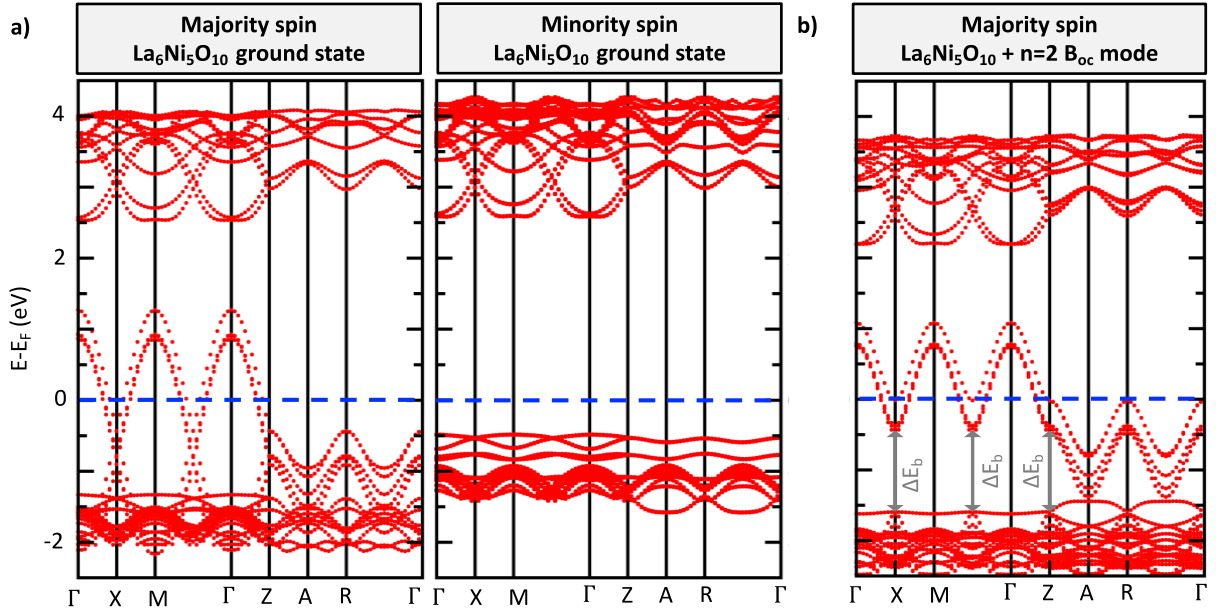

**Figure 8:** Band structure of $La_6Ni_5O_{12}$ ground state (**a**) and with the addition of a finite amplitude Q=0.15 Å/f.u of the n=2 mode (**b**) in the ground state structure. A ferromagnetic order is used.

We then proceed to evaluate the EPC. Since modes are optics and isotropic, we use **equation 9** and obtain an EPC $\lambda$=0.61. This value is reminiscent of $\lambda$=0.51 computed in infinite layered nickelates phases [9] but also with experimental data analysis suggesting an EPC oscillating between 0.56 and 0.61 [43]. Finally, we use the Mc Millan-Allen equation (**equation 11**) for computing the critical temperature. The log average frequency $\omega_{log}$ is estimated as $\omega_{log} = (\omega_2\omega_3)^{1/2}$=67 meV (**equation 12**). Within this approximation and taking typical values of screening parameter μ*=0.1-0.15, we estimate a critical temperature to $T_c$ between 11 and 19 K. These values are close to the experimental $T_c$ of $Nd_6Ni_5O_{12}$ ($T_c$=13 K [6]) and $La_{0.8}Sr_{0.2}NiO_2$ ($T_c$=9-14 K [5,8]), where both systems present an effective $Ni^{1.2+}$ $3d^{8.8}$ configuration.

### IV. **Conclusions**

We explain the electron doping phase diagram of RRP nickelate compounds and show that superconductivity in the appropriately doped nickelate emerges through an EPC associated with charge and bond orderings. This agrees with our previous DFT study on infinite layered nickelates, confirming that there is a common root between all these nickelate superconductors. Although Ni cations present correlation effects inherent to any transition metal element with partly *d* states, the origin of superconductivity is in fact very similar to non-magnetic oxide superconductors such as antimonates and bismuthates once the relevant Ni spin degree of freedom is included in the simulations. This opens the way to reinvestigating superconductivity in oxide superconductors where the role of local spin formation has been previously overlooked.

**Acknowledgements**

This work has received financial support from the CNRS through the MITI interdisciplinary programs under the project SuNi. The work received support from the French Authors acknowledge access granted to HPC resources of Criann through the projects 2020005 and 2007013 and of Cines through the DARI project A0080911453.